\newcommand{\Tr}{\mathop{\mathrm{Tr}}\nolimits} 
\newcommand{\tr}{\mathop{\mathrm{tr}}\nolimits} 
\newcommand{\openone}{\leavevmode\hbox{\small1\normalsize\kern-.33em1}} 
\newcommand{\prim}{\sigma}
\newcommand{\basis}{\theta}
\newcommand{\Gal}[1]{\mathrm{GF}(#1)}

\newcounter{myctr}
\def\myitem{\refstepcounter{myctr}\bibfont\noindent\ifnum\themyctr>9\else\phantom{0}\fi\hangindent17pt\themyctr.\enskip}

\documentclass{ws-ijqi}
\usepackage{bm,amsfonts,amssymb,graphicx,times}

\begin{document}

\markboth{Mu\~{n}oz, Klimov, S\'{a}nchez-Soto and Bj\"{o}rk}
{Coherent states for finite quantum systems}

\catchline{}{}{}{}{}

\title{Discrete coherent states for $n$ qubits}

\author{Carlos Mu\~{n}oz} 
\address{Departamento de F\'{\i}sica,
Universidad de Guadalajara, 44420~Guadalajara,
Jalisco, Mexico}

\author{Andrei B. Klimov} 
\address{Departamento de F\'{\i}sica,
Universidad de Guadalajara, 44420~Guadalajara,
Jalisco, Mexico\\
klimov@cencar.udg.mx}

\author{Luis L. S\'anchez-Soto}
\address{Departamento de \'{O}ptica,
Facultad de F\'{\i}sica, Universidad Complutense,
28040~Madrid, Spain\\
lsanchez@fis.ucm.es}

\author{Gunnar Bj\"{o}rk}
\address{School of Information and
Communication Technology, Royal Institute of Technology (KTH),
Electrum 229, SE-164 40 Kista, Sweden\\
gbjork@kth.se}
\maketitle

\begin{history}
\received{Day Month Year}
\revised{Day Month Year}
\end{history}

\begin{abstract}
  Discrete coherent states for a system of $n$ qubits are introduced
  in terms of eigenstates of the finite Fourier transform. The
  properties of these states are pictured in phase space by resorting
  to the discrete Wigner function.
\end{abstract}


\section{Introduction}

Discrete quantum systems were studied originally by Weyl\cite{Weyl1950}
and Schwinger,\cite{Schwinger1960a} and later by many
authors.\cite{Vourdas2004,Vourdas2007} However, many concepts 
that appear sharp for continuous systems become fuzzy when one
tries to apply them to discrete ones. The reason is that the in
the continuum we have only one harmonic oscillator, while for 
finite systems, there are a number of candidates for that role, 
each one with its own virtues and drawbacks.\cite{Ruzzi2006}

Coherent states constitute an archetypical example of the situation:
for the standard harmonic oscillator they are well understood, and
sensible generalizations have been devised to deal with systems with
more general dynamical groups.\cite{Perelomov1986} However, in spite
of interesting advances,\cite{Galetti1996,Ruzzi2005} the discrete
counterparts are still under heated discussion.

Number states are eigenstates of the Fourier transform,  but they
are not the only ones: coherent states, with their associated
Gaussian wave functions, also are. In our opinion, this subtle, 
yet obvious,  observation has not been taken in due consideration 
in this field. From this perspective, a decisive step was given by 
Mehta,\cite{Mehta1987} who obtained the eigenstates of the 
finite Fourier transform. The purpose of this paper is to further 
explore  this path, showing how we can define physically feasible  
coherent states for $n$ qubits that are also Fourier eigenstates. 

Since the natural arena of these discrete coherent states is the phase
space, we use the discrete Wigner function to picture the
corresponding results. The problem of generalizing the Wigner function
to finite systems has a long story. Using the notions presented in the
comprehensive review~\refcite{Bjork2008}, we construct a Wigner
function for these coherent states and discuss some of their
properties.

\section{Coherent states for $n$ qubits} 

We consider $n$ identical qubits (i.e., $n$ noninteracting spin 1/2
systems). We recall that the Dicke states, belonging to the symmetric
subspace of the representation of SU(2)$^{\otimes n}$, are given by
\begin{equation}
  | n, k \rangle = \sqrt{\frac{k! ( n - k )!}{n!}} \sum_{k} P_{k} 
  ( | 1_{1},  1_{2}, \ldots, 1_{k}, 0_{k+1}, \ldots, 0_{n} \rangle ) \, ,  
  \label{_00}
\end{equation}
where $\{ P_{k} \}$ denotes the complete set of all the possible
permutations of the qubits. These states can be expressed in terms of
the elements of the Galois field $\Gal{2^{n}}$ using the standard
decomposition in the self-dual basis (a short review of the concepts
of finite fields needed in this paper is presented in the appendix)
\begin{equation}
| 1_{1},  1_{2}, \ldots, 1_{k}, 0_{k+1}, \ldots, 0_{n} \rangle  \mapsto 
| 1\sigma_{1} + 1\sigma_{2} + \ldots + 1\sigma_{k} +
0\sigma_{k+1} + \ldots  + 0\sigma_{n} \rangle \, .
\end{equation}
Now, consider a SU(2) coherent state\cite{Perelomov1986}
\begin{equation}
  | \xi \rangle = \frac{1}{( 1+ | \xi |^{2} )^{n/2}}
  \sum_{k=0}^{n}   \sqrt{\frac{n!}{k! ( n - k)!}} \; 
  \xi^{k} | n, k \rangle \, ,
\end{equation}
where the complex number $\xi$ is related with the angular coordinates
$(\vartheta, \varphi)$ on the Bloch sphere by
\begin{equation}
  \label{eq:Bloch}
 \xi = \cot (\vartheta/2) \, e^{- i \varphi} \, .
\end{equation}
Using the previous correspondence, $| \xi \rangle$ can be
recast as
\begin{equation}
  | \xi \rangle = \frac{1}{ ( 1 + | \xi |^{2} )^{n/2}}
  \sum_{\gamma \in \Gal{2^{n}}} \xi^{h (\gamma) } \, | \gamma \rangle \, ,  
  \label{gen_eq}
\end{equation}
where the function $h (\gamma)$, when applied to the field element  
$\gamma =\sum_{k=1}^{n} \gamma_{k}\sigma_{k}$  indicates the number of 
nonzero coefficients $\gamma_{k}$.

In this case, the Fourier operator is 
\begin{equation}
  F = \frac{1}{2^{n/2}} \sum_{\mu , \nu \in \Gal{2^{n}}} 
  \chi ( \mu \nu ) \, | \mu \rangle \langle \nu | \, ,
  \label{_02}
\end{equation}
$\chi$ being an additive character defined in (\ref{Eq: addchardef}).
As it is well known, $F^2 = \openone$, so if we impose that the states 
$| \xi \rangle$ are also eigenstates of $F$ we are lead to
\begin{equation}
  F | \xi \rangle = \pm | \xi \rangle \, .
\end{equation}
This immediately implies (all the spins are pointing in the same direction) 
that there are two SU(2) coherent states (with $\xi_{\pm} = \pm \sqrt{2} - 1$) 
that simultaneously are eigenstates of the Fourier operator: they are 
precisely our candidates to be coherent states for $n$ qubits. In
particular, $| \xi_{\pm} \rangle $ satisfy the following condition
\begin{equation}
  \frac{1}{2^{n/2}} \sum_{\mu , \gamma \in \Gal{2^{n}}} 
  \xi^{h( \gamma )} \chi ( \mu \gamma )  | \mu \rangle =
  \pm \sum_{\mu \in \Gal{2^{n}}} \xi^{h (\mu)} | \mu \rangle \, ,
\end{equation}
or, equivalently,
\begin{equation}
  \frac{1}{2^{n/2}} \sum_{\gamma \in \Gal{2^{n}}}
    \xi^{h ( \gamma)} \chi ( \mu \gamma ) = 
    \pm \xi ^{h (\mu )} \, ,
  \end{equation}
and the minus sign may appear only for odd number of qubits.

Equation (\ref{gen_eq}) is the abstract form of the SU(2) coherent
state. It factorizes in a product of single-qubit states when
represented in the self-dual basis, i.e.,
\begin{equation}
| \xi  \rangle = \frac{1}{( 1 + | \xi |^{2} )^{n/2}}
\sum_{c_{1},\ldots ,c_{n} \in \mathbb{Z}_2}  
\xi^{h\left(\sum_{k=1}^{n} c_{k} \sigma_{k} \right) }
| c_{1} \rangle \ldots | c_{n} \rangle = 
\prod_{j=1}^{n} \frac{(| 0 \rangle + \xi | 1 \rangle )_{j}}
{(1 + | \xi |^{2})^{1/2}} \, ,
\end{equation}
$c_{k}$ being the expansion coefficients of $\gamma$ in that
basis. The operator transforming from the arbitrary basis $\{
\varepsilon_{1}, \varepsilon_{2}, \ldots , \varepsilon_{n} \}$ into 
a factorized form is always a permutation given by
\begin{equation}
P = \sum_{\mu \in \Gal{2^{n}}} ( | \mu_{1}^\prime \rangle  \ldots  
| \mu_{n}^\prime \rangle ) \, ( \langle \mu_{1} | \ldots 
\langle \mu_{n} |) \, ,
\end{equation}
where  
\begin{equation}
\mu = \sum_{i=1}^{n} \mu_{i} \, \varepsilon_{i} = 
\sum_{i=1}^{n} \mu_{i}^\prime \, \sigma_{i} \, ,
\qquad \mu \in \Gal{2^{n}}, 
\quad \mu_{i}, \mu_{i}^\prime \in \mathbb{Z}_2 \, .
\end{equation}

Let us examine the simple yet illustrative example of a two-qubit 
coherent state. In its abstract form it reads as
\begin{equation}
| \xi \rangle = \frac{1}{1 + \xi^{2}}
( | 0 \rangle + \xi | \sigma \rangle + \xi | \sigma^{2} \rangle +
\xi^{2} | \sigma^{3} \rangle ) .
\end{equation}
In the self-dual basis ($\sigma ,\sigma ^{2}$) we have the representation
\begin{equation}
|0\rangle = |00 \rangle =
\left ( 
\begin{array}{c}
0 \\ 
0 \\ 
0 \\ 
1
\end{array}
\right ) , \,
| \sigma \rangle =|10 \rangle = 
\left( 
\begin{array}{c}
0 \\ 
0 \\ 
1 \\ 
0
\end{array}
\right ) , \,
|\sigma^{2} \rangle =| 01 \rangle =
\left ( 
\begin{array}{c}
0 \\ 
1 \\ 
0 \\ 
0
\end{array}
\right ) , \, 
|\sigma^{3} \rangle = |11 \rangle = 
\left ( 
\begin{array}{c}
1 \\ 
0 \\ 
0 \\ 
0
\end{array}
\right ) ,
\end{equation}
in such a way that
\begin{equation}
| \xi \rangle = \frac{1}{1 + \xi^{2}}
\left ( 
\begin{array}{c}
\xi^{2} \\ 
\xi \\ 
\xi \\ 
1
\end{array}
\right ) = 
\frac{1}{\sqrt{1 + \xi^{2}}}
\left ( 
\begin{array}{c}
\xi \\ 
1
\end{array}
\right ) 
\otimes 
\frac{1}{\sqrt{1 + \xi^{2}}}
\left ( 
\begin{array}{c}
\xi \\ 
1
\end{array}
\right ) \, .
\end{equation}
In a non self-dual basis ($\sigma ,\sigma^{3}$) we have 
\begin{equation}
|0 \rangle =|00 \rangle =
\left ( 
\begin{array}{c}
0 \\ 
0 \\ 
0 \\ 
1
\end{array}
\right ) , \, 
| \sigma \rangle = |10 \rangle = 
\left ( 
\begin{array}{c}
0 \\ 
0 \\ 
1 \\ 
0
\end{array}
\right ) ,  \,
| \sigma^{3} \rangle = |01 \rangle =
\left ( 
\begin{array}{c}
0 \\ 
1 \\ 
0 \\ 
0
\end{array}
\right ) , \,
| \sigma^{2} \rangle = |11 \rangle =
\left ( 
\begin{array}{c}
1 \\ 
0 \\ 
0 \\ 
0
\end{array}
\right ) \, ,
\end{equation}
and 
\begin{equation}
| \xi \rangle = \frac{1}{1+\xi^{2}}
\left ( 
\begin{array}{c}
\xi \\ 
\xi^{2} \\ 
\xi \\ 
1
\end{array}
\right ) \, ,
\end{equation}
which cannot be factorized. The transition operator for this 
case is 
\begin{equation}
P = 
\left ( 
\begin{array}{cccc}
0 & 1 & 0 & 0 \\ 
1 & 0 & 0 & 0 \\ 
0 & 0 & 1 & 0 \\ 
0 & 0 & 0 & 1
\end{array}
\right ) \, ,
\end{equation}
and it is nothing but a CNOT gate performing the operation
\begin{equation}
  \label{eq:2}
  |00 \rangle + |01 \rangle  \rightarrow 
  |00 \rangle + |11 \rangle \, .   
\end{equation}

\section{Discrete Wigner function}

To gain further insights into the coherent states $| \xi_{\pm} \rangle$
we proceed to picture them in phase space. To this end, we first
note that, while in the continuous case it is possible to translate
a state by an infinite distance, this is clearly not possible
if the space is finite. To ``prevent'' a state from ``escaping''
the finite phase space it is natural and convenient to use
the field $\Gal{2^n}$ in the representation of the states.

In consequence, we denote by $|\alpha \rangle$, with $\alpha \in
\Gal{2^n}$, an orthonormal basis in the Hilbert space of the system.
Operationally, the elements of the basis can be labeled by powers of a
primitive element, and the basis reads
\begin{equation}
  \{|0 \rangle, \, |\prim \rangle, \ldots, \,|\prim^{2^n-1} = 1 \rangle \} \, .
\end{equation}
These vectors are eigenvectors of the operators $Z_{\beta}$ belonging 
to the generalized Pauli group, whose generators are now defined as
\begin{equation}
  Z_{\beta}  =  \sum_{\alpha \in \Gal{2^n}} \chi ( \alpha \beta ) \,
  | \alpha \rangle \langle \alpha | \, ,
  \qquad \qquad
  X_{\beta} = \sum_{\alpha \in \Gal{2^n}}
  | \alpha + \beta \rangle \langle \alpha | \, , 
  \label{XZgf} \\
\end{equation}
so that
\begin{equation}
  Z_{\alpha} X_{\beta} = \chi ( \alpha \beta ) \, X_{\beta} Z_{\alpha} \, .
\end{equation}

The operators (\ref{XZgf}) can be factorized into tensor products of powers
of single-particle Pauli operators $\sigma_{z}$ and $\sigma_{x}$, whose
expression in the standard basis of the two-dimensional Hilbert space is
\begin{equation}
  \hat{\sigma}_{z} = | 1 \rangle \langle 1 | - |0 \rangle \langle 0 | \, ,
  \qquad \qquad
  \hat{\sigma}_{x} = | 0 \rangle \langle 1 | + | 1 \rangle \langle 0 | \, .
  \label{sigmas}
\end{equation}
This factorization can be carried out by mapping each element of
$\Gal{2^n}$ onto an ordered set of natural numbers.  As we have
already seen, a convenient  choice for this is the self-dual basis, 
since the finite Fourier transform factorizes then into a product of 
single-particle Fourier operators, which leads to
\begin{equation}
  Z_{\alpha}  = \hat{\sigma}_{z}^{a_{1}} \otimes \ldots \otimes
  \hat{\sigma}_{z}^{a_{n}} \, ,
  \qquad \qquad
  X_{\beta} = \hat{\sigma}_{x}^{b_{1}} \otimes \ldots \otimes
  \hat{\sigma}_{x}^{b_{n}} \, , 
\end{equation}
where $(a_{1}, \ldots, a_{n})$ and $(b_{1}, \ldots, b_{n})$ are
the expansion coefficients of $\alpha$ and $\beta$, respectively,
in the self-dual basis.

It was shown that the operators
\begin{equation}
D (\alpha, \beta) = \phi (\alpha, \beta) \, Z_{\alpha} X_{\beta} \, ,  
\label{D_phi}
\end{equation}
where $\phi (\alpha, \beta)$ is a phase, form an operational basis 
in the discrete phase space.\cite{Bjork2008} The unitarity condition
imposes the condition $\phi^{2} ( \alpha, \beta ) = \chi ( - \alpha \beta )$. 
These displacement operators (or phase-point operators in the notation of 
Wootters\cite{Wootters1987}) allows us to introduce a Hermitian kernel
\begin{equation}
  \Delta (\alpha , \beta ) = \frac{1}{2^n} 
  \sum_{\mu , \nu \in \Gal{2^n}}  \chi (\alpha \nu - \beta \mu ) \, 
  D(\mu, \nu ) \, ,  
  \label{NUCLEO}
\end{equation}
in terms of which we can define a well-behaved Wigner function as
\begin{equation}
  W_{\varrho} (\alpha , \beta ) = \Tr [ \varrho \, \Delta (\alpha , \beta ) ] \,  ,  
  \label{WS}
\end{equation}
where $\varrho$ is the density matrix of the system.

\begin{figure}[t]
\centering
\includegraphics[width=0.95\columnwidth]{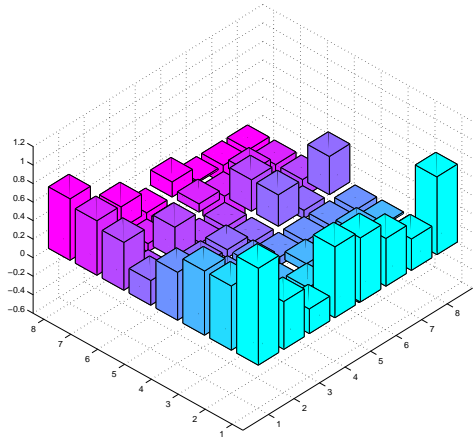}
\vspace*{8pt}
\caption{Wigner function for a coherent state of a system
of three identical qubits.}
\end{figure}

After some calculations, the Wigner function for our coherent states 
turns out to be
\begin{equation}
  W_{|\xi_{+} \rangle} ( \alpha, \beta ) = \frac{1}{2^{n}} 
  \frac{1}{( 1 + | \xi |^{2})^{n}} \sum_{\mu, \nu, \gamma \in \Gal{2^{n}}}
  \xi^{h (\gamma) } \overline{\xi}^{h ( \gamma + \nu) }
  \chi ( \alpha \nu + \beta \mu + \mu \nu + \mu \gamma ) 
  \phi ( \mu , \nu ) \, .
\end{equation}
A plot of this function for the case of three qubits is shown in 
Fig.~1. We also note that the marginal distributions take a very 
simple form: 
\begin{equation}
\sum_{\alpha} W_{| \xi_{+} \rangle} ( \alpha , \beta ) = 
\frac{|\xi^{h ( \beta ) }|^{2}}{( 1 + | \xi |^{2})^{n}}, 
\qquad \qquad
\sum_{\beta } W_{| \xi_{+} \rangle} ( \alpha , \beta ) = 
\frac{|\xi^{h ( \alpha ) }|^{2}}{( 1+ | \xi |^{2})^{n}} \, .
\end{equation}

To conclude we wish to mention that it is also possible to introduce
the notion of squeezing for these states\cite{Marchiolli2007}.  In the
basis of the eigenstates of $Z_{\alpha }$, such an operator has the
following form
\begin{equation}
  S_{\lambda} = \sum_{\kappa \in \Gal{2^{n}}} 
  | \kappa \rangle \langle \lambda \kappa | \, .  
\label{s3}
\end{equation}
The following relations hold
\begin{equation}
  S_{\lambda}^\dagger Z_{\alpha} S_{\lambda} = Z_{\alpha \lambda^{-1}} \, ,
  \qquad \qquad
  S_{\lambda}^\dagger X_{\alpha} S_{\lambda} = X_{\alpha \lambda} \, ,
\end{equation}
so that
\begin{equation}
  \langle \xi_{\pm} | S_{\lambda}^\dagger X_{\alpha} S_{\lambda} | \xi_{\pm} \rangle =
  \langle \xi_{\pm} | S_{\lambda}^\dagger Z_{\alpha \lambda^{2}} S_{\lambda} | 
  \xi_{\pm} \rangle \, .
\end{equation}

\section{Conclusions}

In summary, we have formulated a new sensible approach to 
deal with coherent states for a system of $n$ qubits. 
The associated discrete Wigner function has also been worked out.
Some related problems, as the behavior under time evolution
or the extension to systems of qudits, will be addressed
elsewhere.

\appendix 

\section{Galois fields}
\label{Sec: Galois}

We briefly recall the minimum background of finite fields needed to
proceed through this paper.  The reader interested in more mathematical 
details is referred, e.g., to the excellent monograph by Lidl and 
Niederreiter.\cite{Lidl1986}

A commutative ring is a set $R$ equipped with two binary operations,
called addition and multiplication, such that it is an Abelian group
with respect the addition, and the multiplication is associative. 
Perhaps, the motivating example is the ring of integers $\mathbb{Z}$ 
with the standard sum and multiplication. On the other hand, the 
simplest example of a finite ring is the set $\mathbb{Z}_n$
of integers modulo $n$, which has exactly $n$ elements.

A field $F$ is a commutative ring with division, that is, such that 0
does not equal 1 and all elements of $F$ except 0 have a multiplicative 
inverse (note that 0 and 1 here stand for the identity elements for the 
addition and multiplication, respectively, which may differ from the 
familiar real numbers 0 and 1). Elements of a field form Abelian 
groups with respect to addition and multiplication (in this latter
case, the zero element is excluded).

The characteristic of a finite field is the smallest integer $p$ such 
that
\begin{equation}
p \, 1= \underbrace{1 + 1 + \ldots + 1}_{\mbox{\scriptsize $p$ times}}=0
\end{equation}
and it is always a prime number. Any finite field contains a prime 
subfield $\mathbb{Z}_p$ and has $d = p^n$ elements, where $n$ is a 
natural number. Moreover, the finite field containing $p^{n}$ elements 
is unique and is called the Galois field $\Gal{p^n}$.

Let us denote as $\mathbb{Z}_p[x]$ the ring of polynomials with
coefficients in $\mathbb{Z}_p$. Let $P(x)$ be an irreducible
polynomial of degree $n$ (i.e., one that cannot be factorized over
$\mathbb{Z}_p$). Then, the quotient space $\mathbb{Z}_p[X]/P(x)$
provides an adequate representation of $\Gal{p^n}$. Its elements can 
be written as polynomials that are defined modulo the irreducible 
polynomial $P(x)$. The multiplicative group of $\Gal{p^n}$ is cyclic
and its generator is called a primitive element of the field.

As a simple example of a nonprime field, we consider the 
polynomial $x^{2}+x+1=0$, which is irreducible in $\mathbb{Z}_{2}$.
If $\prim$ is a root of this polynomial, the elements $\{ 0, 1, \prim , 
\prim^{2} = \prim + 1 = \prim^{-1} \} $ form the finite field $\Gal{2^2}$ 
and $\prim$ is a primitive element.

A basic map is the trace
\begin{equation}
 \label{deftr}
 \tr (\alpha ) = \alpha + \alpha^{2} + \ldots +
 \alpha^{p^{n-1}} \, ,
\end{equation}
which satisfies
\begin{equation}
 \label{tracesum}
 \tr ( \alpha + \beta ) =
 \tr ( \alpha ) + \tr ( \beta ) ,
\end{equation}
and  leaves the prime field invariant. In terms of it we define 
the additive characters as
\begin{equation}
 \label{Eq: addchardef}
 \chi (\alpha ) = \exp \left [ \frac{2 \pi i}{p}
  \tr ( \alpha ) \right] ,
\end{equation}
and posses two important properties:
\begin{equation}
 \chi (\alpha + \beta ) =
 \chi (\alpha ) \chi ( \beta ) ,
 \qquad \qquad
 \sum_{\alpha \in
  \Gal{p^n}} \chi ( \alpha \beta ) = p^n
 \delta_{0,\beta} .
 \label{eq:addcharprop}
\end{equation}

Any finite field $\Gal{p^n}$ can be also considered as an
$n$-dimensional linear vector space. Given a basis $\{ \basis_{k} \}$,
($k = 1,\ldots, n$) in this vector space, any field element can be
represented as
\begin{equation}
  \label{mapnum}
  \alpha = \sum_{k=1}^{n} a_{k} \, \basis_{k} ,
\end{equation}
with $a_{k}\in \mathbb{Z}_{p}$. In this way, we map each element of
$\Gal{p^n}$ onto an ordered set of natural numbers $\alpha
\Leftrightarrow (a_{1}, \ldots ,a_{n})$.

Two bases $\{ \basis_{1}, \ldots, \basis_{n} \} $ and $\{
\basis_{1}^\prime, \ldots , \basis_{n}^\prime \} $ are dual when
\begin{equation}
  \tr ( \basis_{k} \basis_{l}^\prime ) =\delta_{k,l}.
\end{equation}
A basis that is dual to itself is called self-dual.

There are several natural bases in $\Gal{p^n}$. One is the polynomial
basis, defined as
\begin{equation}
 \label{polynomial}
 \{1, \prim, \prim^{2}, \ldots, \prim^{n-1} \} ,
\end{equation}
where $\prim $ is a primitive element. An alternative is the normal
basis, constituted of
\begin{equation}
 \label{normal}
 \{\prim, \prim^{p}, \ldots, \prim^{p^{n-1}} \}.
\end{equation}
The choice of the appropriate basis depends on the specific problem 
at hand. For example, in $\Gal{2^2}$ the elements $\{ \prim , \prim^{2}\}$ 
are both roots of the irreducible polynomial. The polynomial basis 
is $\{ 1, \prim \} $ and its dual is $\{ \prim^{2}, 1 \}$, while 
the normal basis $\{ \prim , \prim^{2} \} $ is self-dual.

\vspace*{-6pt}   

\end{document}